\documentclass[11pt,a4paper]{article}
\usepackage{epsf,epsfig,amsfonts,amsgen,amsmath,amstext,amsbsy,amsopn,amsthm}
\usepackage{amssymb}
\usepackage{cite}
\usepackage{enumitem}
\theoremstyle{definition}

\begin{document}
\title{\bf {\Large Two families of  Entanglement-assisted quantum MDS codes from constacyclic codes }}
\date{}

\author{Liangdong Lu$^{a,b,\dag}$, Wenping Ma$^{a}$,  Luobin Guo$^{b}$\\
 a.  State Key Laboratory of Integrated Services Networks,  Xidian University,  \\Xi'an, Shaanxi, 710051,\ China, (email:
$^{\dag}$ kelinglv@163.com, )\\
b. Department of Basic Science,  Air Force Engineering University,
Xi'an,
 Shaanxi,\\ 710051, China}

\maketitle

\begin{abstract} Entanglement-assisted quantum error correcting codes (EAQECCs) can
be derived from arbitrary classical linear codes. However, it is a very difficult task to
determine the number of entangled states required. In this work,
using the method of the decomposition of the defining set of
constacyclic codes, we construct two families of $q$-ary
entanglement-assisted quantum MDS (EAQMDS) codes based on classical
constacyclic MDS codes by exploiting less pre-shared maximally
entangled states. We show that a class  of $q$-ary EAQMDS have
minimum distance upper bound greater than $q$. Some of them have
much larger minimum distance than the known quantum MDS (QMDS) codes
of the same length. Most of these $q$-ary EAQMDS codes are new in
the sense that their parameters are not covered  by the codes
available in the literature.

\medskip

\end{abstract}

\section{\label{sec:level1} Introduction\protect}

Since the significant discovery was published in \cite{shor,cal1},
the theory of quantum error-correcting codes has experienced
tremendous grouth. The most widely studied construction of quantum
error-correcting codes is the stabilizer
formalism\cite{cal2,Ketkar}. It allows standard quantum codes to be
constructed from dual-containing (or self-orthogonal) classical
codes \cite{cal2}. However,  the dual-containing condition  forms  a
barrier in the development of quantum coding theory. In Ref.
\cite{bru06}, Brun {\it et al.} proposed a entanglement-assisted
(EA) stabilizer formalism which shown that non-dual-containing
classical codes can be used to construct EAQECCs if shared
entanglement is available between the sender and receiver.

Let $q$ be a prime power. A $q$-ary  $[[n,k,d;c]]$ EAQECC  that
encodes $k$ information qubits into $n$ channel qubits with the help
of  $c$ pairs of maximally-entangled Bell states  can correct up to
$\lfloor\frac{d-1}{2}\rfloor$  errors, where  $d$ is the minimum
distance of the code. A $q$-ary
$[[n,k,d;c]]$ EAQECC is denoted by $[[n,k,d;c]]_{q}$.  As the same as in classical coding theory, one of the
fundamental tasks in quantum coding theory is to construct quantum
codes with the best possible minimum distance. For EA-quantum
codes, one of the most known bound is EA-Quantum Singleton Bound as follow.

{\bf  Theory 1.1 \cite{bru06,lai}.} ( EA-Quantum Singleton Bound) An
$[[n,k,d;c]]_{q}$ EAQECC satisfies $$n+c-k\geq 2(d-1),$$ where
$0\leq c \leq n-1$.

A code for which equality holds in the  EA-Quantum Singleton Bound is called  EA-quantum
 maximum-distance-separable (EA-QMDS). If $c=0$, then this bound
 is Quantum Singleton Bound, and a code for which equality holds in the bound is called
 quantum maximum-distance-separable (QMDS).
Both QMDS codes  and EAQMDS codes are an important type of quantum
codes. Therefore, constructing QMDS codes and EAQMDS codes have became a
central topic for quantum error-correction codes. Currently, many QMDS codes have been constructed by different
methods\cite{Li3,Chen,He,Jin,Kai1,Kai2,Wang,ZhangT1,ZhangT2}.
In \cite{Ketkar}, the MDS conjecture shown that the length of
maximal-distance-separable (MDS) code cannot exceed $q^{2}+1$. It is shown that except for
some spares lengths $n$, all known $q$-ary quantum MDS codes have
minimum distance less than or equal to $\frac{q+1}{2}$. It is
a very difficult task to construct quantum MDS codes with relatively large
minimum distance. In order to construct some quantum MDS codes with larger minimal distance,
many scholars refer to construct  EA-quantum MDS codes that have
larger minimal distances than the quantum codes with the same length $n$.
One of the most frequently used construction methods is as follows.

 {\bf Proposition 1.2} \cite{bru06,Wil1}.\ \   Let  $\mathcal {C}$$=[n,k,d]_{q^{2}}$
be a classical code over $F_{q^{2}}$ with parity check
matrix $H$. There exists an $[[n,2k-n+c,d;c]]_{q}$ EAQECC,
 where $c=$rank$(HH^{\dagger})$ is the number of maximally entangled states
required and $H^{\dagger}$ is the conjugate
 matrix of $H$  over $F_{q^{2}}$.

In resent years, lots of scholars have constructed many
entanglement-assisted quantum codes with good parameters, see \cite{bru06,
Wil1,lai,Hsi, lai2, lai3,  Fujiwara, Wilde, Lu1, Lu2, guo,Qian1,Qian2,Chen2,lu3,lu4,Galindo,Qian11,Koroglu,Chen3,Liu3,Chen4} .
In \cite{lu4}, we proposed
 the concept about a decomposition of the defining set of constacyclic codes. With the help of this concept,
  we construct some good entanglement-assisted
quantum MDS codes.  In order to discover characters of concept and its applications,
in this paper,   we continue to construct two families of EA-quantum MDS codes
with length $n$ from classical constacyclic codes by the decomposition of the defining set. In \cite{Chen},
it is shown that dual-containing constacylic codes over $F_{q^{2}}$
exist only if the order $r$ is a divisor of $q+1$. So, We pay attention to  constacyclic codes with $r|q+1$
that  can be used to construct
EA-quantum MDS codes.
 More precisely, Our main contribution on new
$q$-ary quantum MDS codes is as follows:

(1)$[[\frac{q^2-1}{2h},\frac{q^2-1}{2h}-2d+3,d;1]]_{q}$, where $q$
is an odd prime power, $h\in \{3,5,7\}$, $2h|q+1$ and
$\frac{q+1}{h}+1\leq d \leq \frac{(q+1)(h+3)}{2h}-1$.

(2)$[[2\lambda(q-1),2\lambda(q-1)-2d+2i,d;2i]]_{q}$, where $q$ be an
odd prime power with $8|q+1$, $\lambda$ is an odd divisor of $q+1$
and $\frac{q-1}{2}(i-1)+4\lambda +1 \leq d \leq
\frac{q-1}{2}+2(i+1)\lambda$;

In construction (1), consumed only one pair of maximally entangled states,
some EA-quantum MDS codes with the
minimal distance upper limit larger than $q$ are obtained.In construction (2),
consumed two pairs of maximally entangled states, some of EA-quantum MDS codes
with the minimal distance are larger than the standard quantum MDS codes in Ref.\cite{Chen}.
Consumed four pairs of maximally entangled states, most of  all EA-quantum MDS codes
with the minimal distance are much larger than $q$.

The paper is organized as follows. In Section 2, basic results about constacyclic codes  and EA-quantum codes are provided.
The concept of a decomposition of the defining set of constacyclic codes is stated.
In Section 3, we give some new classes of EA-quantum MDS
codes. The conclusion is given in Section 4.

\section{Preliminaries}

In this section, we review some basic results on constacyclic codes,
BCH codes, and EAQECCs for the purpose of this paper. For details on BCH codes and constacyclic
codes can be found in standard textbook on coding theory \cite{Macwilliams,Huffman}, and
 for EAQECCs please see Refs.\cite{bru06,Hsi, Wil1, lai, lai2, lai3, Hsieh, Fujiwara, Wilde, Lu1, Lu2}.

Let $p$ be a prime number and $q$ a power of $p$, ie., $q=p^{l}$ for
some $l>0$. We denotes the finite field with $q^{2}$
elements as $F_{q^{2}}$. Given any $\alpha \in F_{q^{2}}$, the conjugation of
$\alpha$ is denoted by $\overline{\alpha}=\alpha^{q}$. For two
vectors $\mathbf{x}=(x_{1},x_{2},\cdots,x_{n})$ and
$\mathbf{y}=(y_{1},y_{2},\cdots,y_{n})\in F_{q^{2}}^{n}$, the
Hermitian inner product is defined as
$(\mathbf{x},\mathbf{y})_{h}=\sum \overline{x_{i}}y_{i}=\overline{x_{1}}y_{1}+\overline{x_{2}}y_{2}+\cdots+\overline{x_{n}}y_{n}.$
For a linear code $\mathcal{C}$ over $F_{q^{2}}$ of length $n$, the
Hermitian dual code $\mathcal{C}^{\bot _{h}}$ is defined as
 $\mathcal{C}^{\bot _{h}}=\{x\in  F_{q^{2}}^{n} | (x,y)_{h}=0, \forall  y $$\in \mathcal{C}\}$.
If $\mathcal{C}^{\bot _{h}}\subseteq \mathcal{C} $, then
$\mathcal{C}$ is called a Hermitian dual containing
code, and $\mathcal{C}^{\bot _{h}}$ is called a Hermitian self-orthogonal code.

We now recall some results about cyclic codes and classical constacyclic codes. For any
vector $(c_{0},c_{1},\cdots,c_{n-1}) $
$\in F_{q^{2}}^{n}$, a
$q^{2}$-ary linear code $\mathcal{C}$ of length $n$ is called $\eta$-constacyclic
if it is invariant
under the $\eta$-constacyclic shift of $F_{q^{2}}^{n}$:
$$
(c_{0}, c_{1}, \cdots , c_{n-1}) \rightarrow (\eta c_{n-1},
c_{0},\cdots, c_{n-2}),$$ where $\eta$ is a nonzero element of
$F_{q^{2}}$. Moreover, $\mathcal{C}$ is called a cyclic code if
$\eta=1$; and $\mathcal{C}$ is  called a negacyclic code if $\eta=-1$.

 For a constacyclic code $\mathcal{C}$, each codeword $c =
(c_{0}, c_{1}, \cdots, c_{n-1})$ is customarily represented in its
polynomial form: $c(x) = c_{0} + c_{1}x + \cdots + c_{n-1}x_{n-1},$
and the code $\mathcal{C}$ is in turn identified with the set of all
polynomial representations of its codewords. For
studying  constacyclic codes, the proper content is the residue class ring
$\mathcal{R}_{n}=\mathbb{F}_{q}[x]/(x^{n}-\eta)$. $xc(x)$ corresponds
to a constacyclic shift of $c(x)$ in the ring $\mathcal{R}_{n}$. As well known,
a linear code $\mathcal{C}$ of length $n$ over $F_{q^{2}}$
is constacyclic if and only if C is an ideal of the quotient ring
$\mathcal{R}_{n}=\mathbb{F}_{q}[x]/(x^{n}-\eta)$. It follows that
$\mathcal{C}$ is generated by monic factors of $(x^{n}-\eta)$, i.e.,
$\mathcal{C}=\langle f(x) \rangle$ and $f(x)|(x^{n}-\eta)$. The $f(x)$
is called the generator polynomial of $\mathcal{C}_{n}$.

Let $\eta \in F_{q^{2}}$ be a primitive $r$th root of unity. Let
 $gcd(n,q)=1$, then there exists a primitive $rn$-th root
of unity $\omega$ in some extension field field of $F_{q^{2}}$ such that
$\omega^{n}=\eta$. Hence, $x^{n}-\eta =\prod ^{n-1}
_{i=0} (x-\omega^{1+ir})$.
Let $\Omega=\{1+ir|0\leq i \leq n-1\}$. For each $j\in \Omega$, let
$C_{j}$ be the $q^{2}$-cyclotomic coset modulo $rn$ containing $j$.
Let $\mathcal{C}$ be an $\eta$-constacyclic code of length $n$ over $F_{q^{2}}$
with generator polynomial $g(x)$. The set  $T=\{j\in\Omega|g(\omega^{j})=0\}$ is
called the defining set of  $\mathcal{C}$.
Let $s$ be an integer with $0\leq s < rn$, the
$q^{2}$-cyclotomic coset modulo $rn$ that contains $s$ is defined by
the set $C_{s}=\{s, sq^{2}, sq^{2\cdot 2}, \cdots, sq^{2(k-1)} \}$
(mod $rn$), where $k$ is the smallest positive integer such that
$xq^{2k}$ $\equiv x$ (mod $rn$).
We can see that the defining set $T$ is a union of some
$q^{2}$-cyclotomic cosets module $rn$ and $dim(\mathcal{C}) =
n-|T|$.

Let $\mathcal {C}$ be a constacyclic code with a defining set $T =
\bigcup \limits_{s \in S} C_{s}$. Denoted $T^{-q}=\{rn-qs | s\in T
\}$, then the  defining set of $\mathcal
{C}$$^{\bot _{h}}$ is $T^{\perp _{h}} =$$ \Omega
$$\backslash T^{-q}$ \cite{Kai2}.
Since there is a striking similarity between cyclic codes and
constacyclic code, we give a correspondence defining of skew aymmetric
and skew asymmetric as follows.
 A cyclotomic coset $C_{s}$ is {\it skew symmetric } if $rn-qs$ mod $rn\in
 C_{s}$; and otherwise is skew asymmetric otherwise. {\it  Skew asymmetric
 cosets}
$C_{s}$ and $C_{rn-qs}$ come in pair, we use $(C_{s},C_{rn-qs})$ to
denote such a pair.

The following results on $q^{2}$-cyclotomic  cosets, dual containing
constacyclic codes are bases of our discussion.

{\bf Lemma 2.1 \cite{Li2,Kai2}.} Let $r$ be a positive divisor of
$q+1$ and $\eta\in F^{*}_{q^{2}}$ be of order $r$. Let $\mathcal{C}$
be a $\eta$-constacyclic code of length $n$ over $F_{q^{2}}$ with
defining set $T$, then $\mathcal {C}$$^{\perp
_{h}}$$\subseteq\mathcal{C}$ if and only if one of the following
holds:

 (1) $T \cap$$T^{-q}=\emptyset$, where $T^{-q}=\{rn-qs \mid s\in
T\}$.

 (2) If $i,j,k\in T$, then
 $C_{i}$ is not a skew asymmetric coset and
($C_{j}$, $C_{k}$) is not  a skew asymmetric cosets pair.

From Lemma 2.1, $\mathcal{C}^{\perp _{h}}\subseteq$
$\mathcal{C}$ can be described by the relationship of its cyclotomic
coset $C_{s}$. However, a defining set $T$ of a non-dual-containing
(or non-self-orthogonal) classical codes is $T \cap$$T^{-q}\neq
\emptyset$.  In order to construct  EA-quantum MDS codes for larger
distance than $q+1$ of code length $n\leq q^2+1$, we recall the
fundamental definition of decomposition of the defining set of
constacyclic codes\cite{lu4}. There are also  other types of
definition for decomposition of the defining set of cyclic codes, negacyclic codes, see
\cite{Li1,Lu1,lu3,Chen2}.

{\bf Definition 2.2\cite{lu4}}  {\it  Let $\eta \in F_{q^{2}}$ be a primitive
$r$th root of unity. Let $ \mathcal {C}$ be a $\eta$-constacyclic
code of length $n$ with defining set $T$. Denote $T_{ss}=T
\cap$$T^{-q}$ and $T_{sas}=T \setminus
$$T_{ss}$, where $T^{-q}=\{rn-qx | x\in T \}$ and $r$ is a factor of $q+1$. $T=T_{ss} \cup
T_{sas}$ is called decomposition of the defining set of
$\mathcal{C}$.}

To  determine $T_{ss}$ and $T_{sas}$, we give the following lemma to
characterize them.

{\bf Lemma 2.3 \cite{Li2}.} Let $gcd(q, n) =
1$, $ord_{rn}$$(q^{2})=m$, $0 \leq x, y$, $z \leq n-1$.

(1) $C_{x}$ is skew symmetric if and only if there is a $t\leq
\lfloor\frac{m}{2}\rfloor$
 such that $x \equiv xq^{2t+1}$(mod n).

(2) If $C_{y}\neq C_{z}$, $(C_{y}, C_{z})$ form a skew asymmetric
pair if and only if there is a $t\leq \lfloor\frac{m}{2}\rfloor$
such that $y \equiv zq^{2t+1}$ (mod n) or $z \equiv yq^{2t+1}$(mod
n).

Using the decomposition of a  defining set  $T$, one can calculate
the number of needed ebits with a algebra method.

 {\bf Lemma 2.4\cite{lu4}.} Let $T$ be a defining set of a constacyclic
code $ \mathcal {C}$, $T=T_{ss}\cup T_{sas}$ be decomposition of
$T$. Using $\mathcal{C}$$^{\perp_{h}}$ as EA stabilizer, the optimal
number of needed  ebits is $c=\mid T_{ss} \mid$.

{\bf Lemma 2.5 \cite{Yang}.} (The BCH bound for Constacyclic Codes)
Let $\mathcal{C}$ be an $\eta$-constacyclic code of length $n$ over
 $F_{q^{2}}$, where $\eta$ is a primitive $r$th root of unity. Let $\omega$
 be a primitive $rn$-th root of unity in an extension field of $F_{q^{2}}$ such that
 $\omega^{n}=\eta$. Assume the generator polynomial of $\mathcal{C}$ has roots that
 include the set $\{\omega^{1+ri}|i_{1}\leq i \leq i_{1}+d-2\}$. Then the minimum distance of $\mathcal{C}$ is at least $d$.

{\bf Lemma 2.6\cite{bru06,Lu1}} Let $\mathcal{C}$ be an $[n,k,d]_{q^{2}}$
constacyclic code with defining set $T$,  and the  decomposition of
$T$ be $T=T_{ss}\cup T_{sas}$. Then $\mathcal{C}$$^{\perp_{h}}$  EA
stabilizes an $q$-ary $[[n,n-2|T|+|T_{ss}|,d \geq \delta ;
|T_{ss}|]]$ EAQECC.

\section{New  EA-quantum MDS Codes}

In this section, we consider $\eta$-constacyclic codes over
$F_{q^{2}}$ of length $n$ to construct EA-quantum codes. To do this,
we give a sufficient condition for a decomposition of the defining
set of $\eta$-constacyclic codes over $F_{q^{2}}$ of length $n$
which do not contain their Hermitian duals. First, we compute
$q^{2}$-cyclotomic cosets modulo $rn$ where $r|q+1$(constacyclic
codes).
\subsection{Lenght
$n=\frac{q^{2}-1}{2h}$ with $h\in \{3,5,7\}$}

Let $h\in \{3,5,7\}$, $q$ be an odd prime power with $2h|(q+1)$.
Suppose $n=\frac{q^{2}-1}{2h}$ and $r=h$. Let $\eta\in F_{q^{2}}$ be
a primitive $r^{th}$ root of unity. Since $rn|q^2-1$ clearly, every
$q^{2}$-cyclotomic coset modulo $rn$ contains exactly one element.

In this subsection, adding one ebit, we construct a new family of a
family of new EA-quantum MDS codes with parameters
$[[\frac{q^{2}-1}{2h},\frac{q^{2}-1}{2h}-2d+3,d;1]]$, where
$\frac{q+1}{2h}+1\leq d \leq \frac{(h+1)(q+1)}{2h}-1$.

{\bf  Lemma 3.1:}   Let $q$ is an odd prime power with $2h|(q+1)$,
$h\in \{3,5,7\}$ and $n=\frac{q^{2}-1}{2h}$. If $\mathcal{C}$ is a
$q^{2}$-ary constacyclic code of length $n$ with define set
$T=\bigcup_{i=\frac{(h-2)(q+1)}{2h}}^{k}\{C_{1+hi}\}$, where
$\frac{(h-2)(q+1)}{2h}\leq k \leq \frac{(2h-1)(q+1)}{2h}-3$, and the
decomposition of a defining set
 $T=T_{ss}\bigcup T_{sas}$, then

\quad(i) $C_{1+(\frac{(h-1)(q+1)}{2h}-1)h}$ is skew symmetric.
\quad(ii) $|T_{ss}|=1$, if $\frac{(h-1)(q+1)}{2h}-1\leq k \leq
\frac{(2h-1)(q+1)}{2h}-3$.

{\bf  Proof:} (i) Since
$1+(\frac{(h-1)(q+1)}{2h}-1)h=\frac{(h-1)(q-1)}{2}$ and
$-[1+(\frac{(h-1)(q+1)}{2h}-1)h]q\equiv
 -\frac{(h-1)(q-1)}{2}q$ $\equiv -[\frac{h-1}{2}(q^{2}-1)-\frac{(h-1)(q-1)}{2}]$
  $\equiv \frac{(h-1)(q-1)}{2}$  $\equiv 1+(\frac{(h-1)(q+1)}{2h}-1)h$ mod $hn$,
  $C_{1+(\frac{(h-1)(q+1)}{2h}-1)h}$ is  skew symmetric.

(ii) Let $T=\bigcup_{i=\frac{(h-3)(q+1)}{2h}}^{k}\{C_{1+hi}\}$,
where $\frac{(h-3)(q+1)}{2h}\leq k \leq q-2$. Since
$C_{1+(\frac{(h-1)(q+1)}{2h}-1)h}$ is skew symmetric, $T_{ss}$
comprises the set $\{C_{1+(\frac{(h-1)(q+1)}{2h}-1)h}\}$ at least.
 According to the concept about a  decomposition of the  defining set $T$,
 one obtain that $T_{sas}=T\backslash T_{ss}$.
 In order to testify $|T_{ss}| =1$ if $\frac{(h-1)(q+1)}{2h}-1\leq i \leq q-2$, from Definition 2.2 and Lemma 2.3,
 we need  to testify that there is no skew symmetric cyclotomic
 coset, and any two cyclotomic
 coset do not form a  skew asymmetric pair in  $T_{sas}$.

Let $I=\{1+hi| \frac{(h-3)(q+1)}{2h}\leq i \leq q-2\}\setminus
(1+(\frac{(h-1)(q+1)}{2h}-1)h)$ and $r=h$. Only we need  to testy
that for $\forall x\in I$, $-qx$ (mod $rn$)$\not \in I$ and
$T_{ss}=\{C_{1+(\frac{(h-1)(q+1)}{2h}-1)h}\}$. That implies that if
$x,y\in I$,  $C_{x}$ is not a skew symmetric
cyclotomic  coset, and any $C_{x},C_{y}$ do not form a skew
asymmetric pair if and only if $x+yq\not\equiv0$ mod $rn$.

Divide $I$ into three parts
$I_{1}=[1+\frac{(h-3)(q+1)}{2h}h,1+(\frac{(h-1)(q+1)}{2h}-2)h]$,
$I_{2}=[1+(\frac{(h-1)(q+1)}{2h})h,1+(\frac{(h+1)(q+1)}{2h}-2)h]$
and $I_{3}=[1+(\frac{(h+1)(q+1)}{2h}-1)h,1+(q-2)h]$. $rn=hn=q^2-1$.
 If $ x,y\in I_{1}$, then
$\frac{(h-3)}{2}(q^{2}-1)+(h-2)(q+1)=(1+\frac{(h-3)(q+1)}{2h}h)(q+1)\leq
x+yq \leq
(1+(\frac{(h-1)(q+1)}{2h}-2)h)(q+1)=\frac{(h-1)}{2}(q^{2}-1)-h(q+1)$.
Since $h\in\{3,5,7\}$,  $\frac{(h-3)}{2}=\{0,1,2\}$, and
$\frac{(h-1)}{2}=\{1,2,3\}$. Hence, for $ x,y\in I_{1}$,  $ln< x+yq
< (l+1)n$, where $0\leq l\leq2$ is a integer.

 If $ x,y\in I_{2}$, then
$\frac{h-1}{2}(q^{2}-1)+h(q+1)=(1+(\frac{(h-1)(q+1)}{2h})h)(q+1)\leq
x+yq \leq
(1+(\frac{(h+1)(q+1)}{2h}-2)h)(q+1)=\frac{h+1}{2}(q^{2}-1)-(h-2)(q+1)$.
Since $h\in\{3,5,7\}$,  $\frac{(h-1)}{2}=\{1,2,3\}$, and
$\frac{(h+1)}{2}=\{2,3,4\}$. Hence, for $ x,y\in I_{2}$,  $ln< x+yq
< (l+1)n$, where $1\leq l\leq 3$ is a integer.

For $h=3$, if $ x,y\in I_{3}$, then
$2n<\frac{h+1}{2}(q^{2}-1)+2(q+1)=(1+(\frac{(h+1)(q+1)}{2h}-1)h)(q+1)\leq
x+yq \leq (1+(q-2)h)(q+1)=h(q^{2}-1)-(h-1)(q+1)<3n$.

For $h=5$,  Divide $I_{3}$ into two parts
$I_{3}^{'}=[1+(\frac{(h+1)(q+1)}{2h}-1)h,1+(\frac{(h+3)(q+1)}{2h}-2)h]$
and $I_{3}^{''}=[1+(\frac{(h+3)(q+1)}{2h}-1)h,1+(q-2)h]$.

if $ x,y\in I_{3}^{'}$, then
$3n<\frac{h+1}{2}(q^{2}-1)+2(q+1)=(1+(\frac{(h+1)(q+1)}{2h}-1)h)(q+1)\leq
x+yq \leq
(1+(\frac{(h+3)(q+1)}{2h}-2)h)(q+1)=\frac{h+3}{2}(q^{2}-1)-(h-2)(q+1)<4n$;
if $ x,y\in I_{3}^{''}$, $4n<(1+(\frac{(h+3)(q+1)}{2h}-2)h)(q+1)\leq
x+yq \leq (1+(q-2)h)(q+1)=h(q^{2}-1)-(h-1)(q+1)<5n$.

For $h=7$,  Divide $I_{3}$ into two parts
$I_{3}^{a}=[1+(\frac{(h+1)(q+1)}{2h}-1)h,1+(\frac{(h+3)(q+1)}{2h}-2)h]$,
$I_{3}^{b}=[1+(\frac{(h+3)(q+1)}{2h}-1)h,1+(\frac{(h+5)(q+1)}{2h}-2)h]$
and $I_{3}^{c}=[1+(\frac{(h+5)(q+1)}{2h}-1)h,1+(q-2)h]$.

If $ x,y\in I_{3}^{a}$,
$4n<\frac{h+1}{2}(q^{2}-1)+2(q+1)=(1+(\frac{(h+1)(q+1)}{2h}-1)h)h)(q+1)\leq
x+yq \leq
(1+(\frac{(h+3)(q+1)}{2h}-2)h)(q+1)=\frac{h+3}{2}(q^{2}-1)-(h-2)(q+1)<5n$;
if $ x,y\in I_{3}^{b}$, $5n<(1+(\frac{(h+3)(q+1)}{2h}-2)h)(q+1)\leq
x+yq \leq
(1+(\frac{(h+5)(q+1)}{2h}-2)h)(q+1)=\frac{h+5}{2}(q^{2}-1)-(h-2)(q+1)<6n$.
if $ x,y\in I_{3}^{c}$, $6n<(1+(\frac{(h+3)(q+1)}{2h}-1)h)(q+1)\leq
x+yq \leq (1+(q-2)h)(q+1)=h(q^{2}-1)-(h-1)(q+1)<7n$.

Hence, there is no skew symmetric cyclotomic cosets, and any two
cyclotomic  coset do not form a skew asymmetric pair in $T\setminus
\{C_{1+(\frac{(h-1)(q+1)}{2h}-1)h}\}$. That implies that
$T_{ss}=\{C_{1+(\frac{(h-1)(q+1)}{2h}-1)h}\}$ and $|T_{ss}|=1$ for
$\frac{(h-1)(q+1)}{2h}-1\leq i \leq q-2$, when the defining set
$T=\bigcup_{i=\frac{(h-3)(q+1)}{2h}}^{k}\{C_{1+hi}\}$,
where $\frac{(h-3)(q+1)}{2h}\leq k \leq q-2$.\\

{\bf  Theory 3.2:}  Let $q$ is an odd prime power with $2h|(q+1)$,
$h\in \{3,5,7\}$ and $n=\frac{q^{2}-1}{2h}$.
 Then there exists a q-ary
$[[\frac{q^{2}-1}{2h},\frac{q^{2}-1}{2h}-2d+3,d;1]]$ EAQMDS, where
$\frac{q+1}{2h}+1\leq d \leq \frac{(h+1)(q+1)}{2h}-1$.

{\bf  Proof:} Consider the constacyclic codes over $F_{q^{2}}$ of
length $n=\frac{q^2-1}{h}$ with defining set
$T=\bigcup_{i=\frac{(h-3)(q+1)}{2h}}^{k}\{C_{1+hi}\}$, where
$\frac{(h-3)(q+1)}{2h}\leq k \leq q-2$, $h\in \{3,5,7\}$ and
$h|(q+1)$ for $q$ be an odd prime power. By Lemma 3.1, there is
$c=|T_{ss}|=1$ if $\frac{(h-1)(q+1)}{2h}-1\leq k \leq q-2$. Since
every $q^{2}$-cyclotomic coset has one element which must be odd
number, we can obtain that $T$ consists of
$(i-\frac{(h-3)(q+1)}{2h}+1)$ integers
$\{1+(\frac{(h-3)(q+1)}{2h})h,1+(\frac{(h-3)(q+1)}{2h}+1)h,
1+(\frac{(h-3)(q+1)}{2h}+2)h\cdots,1+kh\}$. It implies that
$\mathcal{C}$ has minimum distance at least
$(i-\frac{(h-3)(q+1)}{2h}+2)$. Hence, $\mathcal{C}$ is a $q^{2}$-ary
constacyclic code with parameters
$[n,n-2(i-\frac{(h-3)(q+1)}{2h}+1)+1,\geq
(i-\frac{(h-3)(q+1)}{2h}+2)]$. Combining Lemma 2.6 with EA-quantum
Singleton bound, we can obtain a EA-quantum MDS code with parameters
$[[\frac{q^{2}-1}{2h},\frac{q^{2}-1}{2h}-2d+3,d;1]]_{q}$, where
$\frac{q+1}{2h}+1\leq d \leq \frac{(h+1)(q+1)}{2h}-1$.(See Table 1)\\

\begin{center}
Table 1 EAQMDS codes with $n=\frac{q^{2}-1}{2h}$ \\
\begin{tabular}{lllllllllll}
  \hline
  q          & h          &$[[n,k,d;1]]_{q}$                  &d                               \\
\hline
 11          & 3          &$[[20,23-2d,d;1]]_{11}$              &$3\leq d \leq 7$                  \\

 23          & 3          &$[[88,91-2d,d;1]]_{23}$               &$5\leq d \leq 15$               \\

 19           &5          &$[[36,39-2d,d;1]]_{19}$                &$3\leq d \leq 11$             \\

 29           &5          &$[[84,87-2d,d;1]]_{29}$                &$4\leq d \leq 17$         \\

 13           &7          &$[[12,15-2d,d;1]]_{13}$                &$2\leq d \leq 7$             \\

 41           &7          &$[[120,123-2d,d;1]]_{41}$                &$4\leq d \leq 23$         \\

 \hline
  \end{tabular}
\end{center}

\subsection{Lenght
$n=2\lambda(q-1)$ with $8|(q+1)$ and $\lambda|(q+1)$ }

Let $q$ be an odd prime power with $8|(q+1)$. Let $\lambda$ be an
odd divisor of $q+1$, $n=2\lambda(q-1)$ and
$r=\frac{q+1}{2\lambda}$. Clearly, $q\geq 7$ and $r\geq 4$. Now, we
use $\eta$-constacyclic codes over $F_{q^{2}}$ of length $n$ to
construct $q$-ary EA-quantum MDS codes of length $n$, where $\eta\in
F_{q^{2}}$ is a primitive $r^{th}$ root of unity.

Let $\mathcal{C}$ be a $\eta$-constacyclic code of length $n$ over
$F_{q^{2}}$ with defining set
$$T=\{1+ri| -(4\lambda-1)\leq i \leq \frac{q-1}{2}+2\lambda-1 \}.$$
Since $2\lambda t=q+1$ and $q\geq 7$,  one can obtain that $0<1+r(\frac{q-1}{2}+2\lambda-1)<\frac{q^2-1}{2}$ and
$-\frac{q^2-1}{2}<1-r(4\lambda-1)<0$.

In \cite{Chen}, if the defining set $T=\{1+ri| -(2t-1)\leq i \leq
4t-2 \}$, Chen et al. have constructed quantum MDS codes with
parameters $[[2\lambda(q-1),2\lambda(q-1)-2d+2,d]]_{q}$ , where
$2\leq d\leq 6\lambda-1$. However, it is very hard to enlarge the
minimal distance $d$ for this type quantum MDS codes. In this
subsection, in order to enlarge the minimal distance, adding few
ebit, we construct a new family of a family of new EA-quantum MDS
codes.

{\bf  Lemma 3.3:}   Let $q$ be an odd prime power with $8|(q+1)$ and
$n=2\lambda(q-1)$ where $\lambda$ is an odd divisor of $q-1$. If
$\mathcal{C}$ is a $q^{2}$-ary negacyclic code of length $n$ with
define set $T=\{1+ri| -(4\lambda-1)\leq i \leq \frac{q-1}{2}+2\lambda-1 \}$
 and the decomposition of a
defining set
 $T=T_{ss}\bigcup T_{sas}$, then

\quad(i)$(C_{1},C_{-q})$ and
 $(C_{1+r(\frac{q-1}{2})},C_{1+(r-4)(\frac{q-1}{2})+q-3})$
 form
skew asymmetric cosets pairs, respectively.

\quad(ii)  $$|T_{ss}| = \left\{
\begin{array}{lll}
2,                 &\mbox {if $-(4\lambda-1)\leq i \leq \frac{q-1}{2}-1$;}\\
4,               &\mbox {if $\frac{q-1}{2} \leq i \leq \frac{q-1}{2}+2\lambda-1$.}\\
 \end{array}
\right. $$

{\bf  Proof:}(i) It is clear that $(C_{1},C_{-q})$ forms a skew asymmetric cosets pair. Since $rn=r2\lambda(q-1)=q^2-1$,
 $(1+r(\frac{q-1}{2}))q=q+\frac{r}{2}(q^{2}-q)$  $\equiv 1+\frac{r}{2}(q-1)-2(q-1)+q-3$    $\equiv 1+(r-4)(\frac{q-1}{2})+q-3$ mod $(q^2-1)$.

(ii) Let $T=\{1+ri| -(4\lambda-1)\leq i \leq \frac{q-1}{2}+2\lambda-1 \}$. Since
$(C_{1},C_{-q})$ and
 $(C_{1+r(\frac{q-1}{2})}, $  $ C_{1+(r-4)(\frac{q-1}{2})+q-3})$ form
skew asymmetric cosets pairs, respectively, $T_{ss}$
comprises the set $\{ $ $C_{1},C_{-q},$ $C_{1+r(\frac{q-1}{2})},$ $C_{1+(r-4)(\frac{q-1}{2})+q-3}\}$ at least.
 According to the concept about a  decomposition of the  defining set $T$,
 one obtain that $T_{sas}=T\backslash T_{ss}$.
 In order to testify $|T_{ss}|=2$ if $-(4\lambda-1)\leq i \leq \frac{q-1}{2}-1$ and $|T_{ss}|=4$ if $\frac{q-1}{2} \leq i \leq \frac{q-1}{2}+2\lambda-1$, from Definition 2.2 and Lemma 2.3,
 we need  to testify that there is no skew symmetric cyclotomic
 coset, and any two cyclotomic
 coset do not form a  skew asymmetric pair in  $T_{sas}$.

Let $I=\{1+ri| -(4\lambda-1)\leq i \leq \frac{q-1}{2}+2\lambda-1\}\setminus
(1,-q,1+r(\frac{q-1}{2}),1+(r-4)(\frac{q-1}{2})+q-3)$ and $r=\frac{q+1}{2\lambda}$. Only we need  to testy
that for $\forall x\in I$, $-qx$ (mod $rn$)$\not \in I$ and
$T_{ss}=\{C_{1},C_{-q},C_{1+r(\frac{q-1}{2})},C_{1+(r-4)(\frac{q-1}{2})+q-3}\}$. That implies that if
$x,y\in I$, from  Lemma 2.3, $C_{x}$ is not a skew symmetric
cyclotomic  coset, and any $C_{x},C_{y}$ do not form a skew
asymmetric pair if and only if $x+yq\not\equiv0$ mod $rn$.

Divide $I$ into $(2+\frac{r}{2})$ parts
$I_{1}=[1-r(4\lambda-1),1-r(2\lambda-1)]$,
$I_{2}=[1-r2\lambda,-1]$, $I_{3}=[1+r,1+r(2\lambda-1)]$,
$I_{4}=[1+r(2\lambda),1+r(4\lambda-1)]$, $\cdots ,$ $I_{(2+\frac{r}{2}-1)}=[1+r(\frac{q-1}{2}-2\lambda),1+r(\frac{q-1}{2}-1)]$ and $I_{(2+\frac{r}{2})}=[1+r(\frac{q-1}{2}+1),1+r(\frac{q-1}{2}+2\lambda-1)]$. $rn=2r\lambda(q-1)=q^2-1$.
 If $ x,y\in I_{1}$, then
$-2n< -2(q^2-1)+(r-3)(q+1)=(1-r(4\lambda-1))(q+1)\leq$
$x+yq \leq
(1-r(2\lambda-1))(q+1)=-(q^2-1)-(r-1)(q+1)<-n$;  If $ x,y\in I_{2}$, then
$-n< -(q^2-1)+(q+1)=(1-r2\lambda)(q+1)\leq$
$x+yq \leq
-(q+1)<0$;  If $ x,y\in I_{3}$, then
$q+1< (1+r)(q+1)\leq$
$x+yq \leq
(1+r(2\lambda-1))(q+1)<n$. Using the same method, if $ x,y\in I_{(2+\frac{r}{2})}$, then
$\frac{r}{2}n< (1+r(\frac{q-1}{2}+1))(q+1)\leq$
$x+yq \leq
(1+r(\frac{q-1}{2}+2\lambda-1))(q+1)<(\frac{r}{2}+1)n$.

Hence, there is no skew symmetric cyclotomic cosets, and any two
cyclotomic  coset do not form a skew asymmetric pair in $T\setminus
\{ $ $C_{1},C_{-q},$ $C_{1+r(\frac{q-1}{2})},$ $C_{1+(r-4)(\frac{q-1}{2})+q-3}\}$ .
That implies that  $|T_{ss}|=2$ if $-(4\lambda-1)\leq i \leq \frac{q-1}{2}-1$ and $|T_{ss}|=4$ if $\frac{q-1}{2} \leq i \leq \frac{q-1}{2}+2\lambda-1$.\\

 {\bf  Theory 3.4:}  Let $q$ be an odd prime power with $8|(q+1)$ and
$n=2\lambda(q-1)$ where $\lambda$ is an odd divisor of $q-1$.
 then there exists a q-ary
$[[2\lambda(q+1),2\lambda(q+1)-2d+2i,d;2i]]$ EAQMDS, where $1\leq i\leq 2$ and $\frac{q-1}{2}(i-1)+4\lambda +1 \leq d \leq
\frac{q-1}{2}+2(i+1)\lambda$.

{\bf  Proof:} Consider the constacyclic codes over $F_{q^{2}}$ of
length $n=2\lambda(q-1)$ with defining set
$T=\bigcup_{i=-(4\lambda-1)}^{k}\{C_{1+ri}\}$, where
$-(4\lambda-1)\leq k \leq \frac{q-1}{2}+2\lambda-1$  and
$8|(q+1)$ for $q$ be an odd prime power. By Lemma 3.3, there is
$c=|T_{ss}|=2$ if $-(4\lambda-1)\leq k \leq \frac{q-1}{2}-1$ and $c=|T_{ss}|=4$ if $\frac{q-1}{2} \leq k \leq \frac{q-1}{2}+2\lambda-1$. Since
every $q^{2}$-cyclotomic coset has one element which must be odd
number, we can obtain that $T$ consists of
$k+(4\lambda-1)+1$ integers
$\{1+r(-(4\lambda-1)),1+r(-(4\lambda-1)+1),
1+r(-(4\lambda-1)+2),\cdots,1+rk\}$. It implies that
$\mathcal{C}$ has minimum distance at least
$k+4\lambda+1$. Combining Lemma 2.6 with EA-quantum
Singleton bound, we can obtain a EA-quantum MDS code with parameters
$[[2\lambda(q+1),2\lambda(q+1)-2d+2i,d;2i]]$ EAQMDS, where $1\leq i\leq 2$ and $\frac{q-1}{2}(i-1)+4\lambda +1 \leq d \leq
\frac{q-1}{2}+2(i+1)\lambda$.(See Table 2)

\begin{center}
Table 2 EAQMDS codes with $n=2\lambda(q-1)$ \\
\begin{tabular}{lllllllllll}
  \hline
  q          & $\lambda$        &$[[n,k,d;c]]_{q}$                  &d                               \\
\hline
 23           & 3          &$[[132,136-2d,d;2]]_{23}$              &$13\leq d \leq 23$                  \\

              &            &$[[132,138-2d,d;4]]_{23}$              &$24\leq d \leq 29$                  \\

 47           &3          &$[[276,280-2d,d;2]]_{47}$              &$13\leq d \leq 35$             \\

              &           &$[[276,282-2d,d;4]]_{47}$              &$36\leq d \leq 41$         \\

 79           &5          &$[[780,784-2d,d;2]]_{79}$                &$21\leq d \leq 59$             \\

              &           &$[[780,786-2d,d;4]]_{79}$                &$60\leq d \leq 69$         \\

 87           &11          &$[[1892,1896-2d,d;2]]_{87}$                &$45\leq d \leq 87$             \\

              &           &$[[1892,1898-2d,d;4]]_{87}$                &$88\leq d \leq 109$         \\

103           &13          &$[[2652,2656-2d,d;2]]_{103}$                &$53\leq d \leq 103$             \\

              &           &$[[2652,2658-2d,d;4]]_{103}$                &$104\leq d \leq 129$         \\

 \hline
  \end{tabular}
\end{center}

\section{ SUMMARY}

In this paper, based on classical
constacyclic MDS codes with a concept about a decomposition of the defining
set of constacyclic codes, we construct two families of $q$-ary
entanglement-assisted quantum MDS (EAQMDS) codes by exploiting less pre-shared maximally
entangled states.
In Table 3, we list the $q$-ary entanglement-assisted quantum MDS
codes constructed in this paper. By consuming two pre-shared
maximally entangled states, we obtain the EA-quantum MDS codes of
$2\lambda(q-1)$ with the minimal distance upper limit greater than $q$ (odd).
These EA-quantum MDS codes are improved the
parameters of codes in Ref.\cite{Chen}. Moreover, consuming only one pair of maximally
entangled states, we obtain a family of EA-quantum MDS codes of
$\frac{q^2-1}{2h}$ with the minimal distance larger than the standard
quantum MDS codes in Ref.\cite{Kai2}.

Comparing the
parameters with $q$-ary EA-quantum MDS codes, we find that these
quantum MDS  codes are new in the sense that their parameters are
not covered by the codes available in the literature.\\

\begin{center}
Table 3 New parameters of EAQMDS codes \\
\begin{tabular}{lllllllllllllll}
  \hline
  Class                    & length                      &$[[n,k,d;c]]_{q}$                &Distance  \\
\hline
 1                        &$n=\frac{q^{2}-1}{2h}$        &$[[\frac{q^2-1}{2h},\frac{q^2-1}{2h}-2d+3,d;1]]_{q}$    &$\frac{q+1}{2h}+1\leq d \leq \frac{(h+1)(q+1)}{2h}-1$\\
                          &$h\in \{3,5,7\}$   & &                \\

2                        &$n=2\lambda(q-1)$             &$[[n,n-2d+2i,d;2i]]_{q}$      &$\frac{q-1}{2}(i-1)+4\lambda +1 \leq d \leq
\frac{q-1}{2}+$\\
                           &$\lambda|q-1$ odd,           &$1\leq i \leq 2$     &$2(i+1)\lambda$    \\
                        &$8|(q+1)$,                      & &    \\

  \hline
  \end{tabular}
\end{center}

\section*{Acknowledgment}
This work is supported by the National Natural
Science Foundation of China under Grant No.11801564,the National Key R\&D Program of China under Grant No. 2017YFB0802400, the National Natural Science Foundation of China under grant No. 61373171, 111 Project under grant No.B08038.

\bibliographystyle{amsplain}

\end{document}